\documentstyle[12pt]{article}   
\begin{document}

\title{{\small\centerline{January 1995 \hfill IIA-NAPP-95-1}}
\medskip
{\bf Constraints on cosmic charge asymmetry and neutrino
charge from the  microwave background}}

\author{\bf Sujan Sengupta\thanks{e-mail: sujan@iiap.ernet.in}
and Palash B. Pal\thanks{e-mail: pbpal@iiap.ernet.in}\\
\normalsize \em Indian Institute of Astrophysics, Bangalore 560034,
             INDIA}

\date{}
\maketitle

\begin{abstract} \normalsize\noindent
Considering the observed anisotropy in cosmic microwave background radiation
($\Delta T/T \leq 10^{-4}$) an upper limit on the electric charge asymmetry
over a cosmological scale is found which is several orders more stringent
than those found earlier. The same argument constrains the
charge of massless, degenerate neutrinos.
\end{abstract}

PACS numbers : 98.80.-k, 98.70.Vc, 95.85.Ry
One possibility which has excited particular interest over the
years is the question of whether matter in particular and the
Universe in general have absolute charge symmetry. Interest in
the subject was aroused following the suggestion by Bondi and
Littleton \cite{1} that if the electron and proton charge
differed by a part in $10^{18}$ it would account for the
expansion of the Universe. This was followed by the experiment
of Hillas and Cranshaw \cite{2}. Afterwards the work gained
general acceptance by the scientific community and there were
several experiments (see for example Ref.~\cite{3}) to detect
charge asymmetry in matter. All these experiments put definite
upper limits on the charge asymmetry although they are much lower
than that required by Bondi and Littleton. However, since the
strength of the electromagnetic interaction is $10^{39}$ times
stronger than the gravitational interaction any small asymmetry
could have a profound consequence. Recently it has been
suggested \cite{4} that a net charge asymmetry could generate
chaotic magnetic field on astronomically interesting scales.

In this context we show in this letter that a slight excess in
electric charge over a cosmic scale can have severe consequence
on the observed isotropy in the cosmic microwave background
radiation and hence put a limit on the net charge asymmetry of
the Universe which is several orders more stringent than the
existing limits. Following the same arguments a stringent upper
limit on the electric charge of massless, degenerate neutrinos
is obtained.

Let $n_{Q}$ be the present excess number density of particles
with charge $e$ over the particles of opposite charge.  Then
$n_{Q}e$ is the present charge density which gives rise to the
charge asymmetry over cosmic scales. Consider now a sphere of
significantly large radius $R$.  The electric forces will
generate peculiar velocities in the particles in very much the
same way the gravitational density perturbations do. Thus, just
as the gravitational density perturbations induce temperature
fluctuations in the microwave background (the Sachs-Wolfe effect
\cite{SW}), the electrical forces will induce the same. To
estimate the effect, we first note that the electric self-energy
of the particles in this sphere can be written as
	\begin{eqnarray}
{\cal E}_R^{(\rm el)} = \frac{16\pi^2}{15} n_Q^2 \alpha R^5 \,,
	\end{eqnarray}
where $\alpha=1/137$ is the fine structure constant.
On the other hand, the gravitational self-energy of the same
sphere due to the uniform background mass density $\rho$ is
given by
	\begin{eqnarray}
{\cal E}_R^{(\rm gr)} =\frac{16\pi^2}{15} G \rho^2 R^5 \,.
	\end{eqnarray}
The ratio of these two gives the temperature fluctuation in the
background radiation:
	\begin{eqnarray}
{\Delta T \over T} \approx
{{\cal E}_R^{(\rm el)} \over {\cal E}_R^{(\rm gr)}}  =
{n_Q^2 \alpha \over G\rho^2} \,.
	\end{eqnarray}
This provides an upper limit on $n_{Q}$ given by
	\begin{eqnarray}
n_{Q} \leq \rho  \sqrt{\frac{G}{\alpha} \times \left(
\frac{\Delta T}{T} \right)_{\rm obs}} \,.
	\end{eqnarray}
Considering the universe to have critical density given by
	\begin{eqnarray}
\rho_c = 10^4 h_0^2 \; {\rm eV/cm}^3 \,,
	\end{eqnarray}
where $h_0$ is the Hubble parameter in units of $100 {\rm
km\,s^{-1}\,Mpc^{-1}}$,  we thus obtain
	\begin{eqnarray}
n_Q \leq 9.6 \times 10^{-26} h_0^2 \; {\rm cm}^{-3} \,,
\label{nq}
	\end{eqnarray}
using $\Delta T/T \leq 10^{-4}$~\cite{5,6}. Equivalently, one
can write
	\begin{eqnarray}
{n_Q \over s} \leq 3.2 \times 10^{-29} h_0^2 \,,
	\end{eqnarray}
where $s$ is the entropy density of the microwave background
radiation, whose value at the present epoch is
$2970\;{\rm cm}^{-3}$ for $T_0=2.75^{\circ}$\,K.

The above limit is several orders more stringent than the
existing limit $n_Q/s\leq 10^{-27}$ \cite{1}. It is comparable
with the stringent bound on the net excess of charge per baryon
constrained to be below $10^{-30}$ from arguments involving the
isotropy of cosmic rays, i.e., anisotropy induced by electric
field \cite{7}. In these arguments for isotropy of high energy
cosmic rays of energy $E$, one requires the product of the
coherence length and the field to be less than $E$ giving the
limit of $10^{-30}$.

We now turn our attention to the bounds on the electric charge
of the neutrino.  Considering the bunching in time of the
neutrinos detected from SN1987A in LMC, Barbiellini and Cocconi
\cite{8} found the upper limit of the neutrino electric charge
to be $10^{-17}e$. This limit is substantially smaller than the
limit $10^{-13}e$ obtained from arguments on solar energy losses
\cite{9}. Assuming the bulk of the dark matter in galactic halos
and clusters constituted by non-zero rest mass neutrinos a more
stringent limit ($10^{-26}e$) is found \cite{10}. Considering
the galactic magnetic field $B_{G}\simeq 10^{-6}$\,G at a cosmic
scale $R=10$\,kpc and the virial velocity of neutrinos in the
galactic magnetic field to be 300 km\,s$^{-1}$ the same author
found the limit of the neutrino electric charge to be $2\times
10^{-32}e$.  Although this is a very stringent limit, its
derivation does not have any strong observational support as it
depends on the hot dark matter hypothesis.  Further, the
numerical value of the charge is highly sensitive to the rest
mass of neutrino which is rather uncertain (only the electron
neutrino mass is considered in this work without taking into
account the opposite charge of antineutrinos).

Our argument from temperature fluctuations can be applied to
constrain the electric charge of massless neutrinos if they are
degenerate.   Let $\mu_i$ be the chemical potential of the
neutrinos where $i$ implies electron, muon and tau neutrinos. In
the degenerate limit ($|\mu_{i}| \gg T_{\nu}$) the massless
neutrinos contribute an amount \cite{11}
	\begin{eqnarray}
\rho_{\nu+\overline{\nu}} \simeq \frac{1}{8\pi^2}
[\sum_{i=1}^3 \mu_i^4]
	\end{eqnarray}
to the present energy density of the Universe. Requiring
$\rho_{\nu+\overline{\nu}} \leq\rho_{c}$, we obtain
	\begin{eqnarray}
\left( \sum_{i=1}^3 \mu^{4}_{i} \right)^{1/4} \leq 8.9\times
10^{-3} \sqrt{h_0} \;{\rm eV}.
	\end{eqnarray}
Assuming all flavors of neutrinos are equally degenerate, this gives
	\begin{eqnarray}
\frac{|\mu_{1}|}{T_\nu} = \frac{|\mu_{2}|}{T_\nu} =
\frac{|\mu_{3}|}{T_\nu} \leq 53 \sqrt{h_0}\,,
\label{50}
	\end{eqnarray}
using $T_{\nu}=1.96^{\circ}$\,K.

The net lepton number density contributed by neutrinos is given
by~\cite{KT}
	\begin{eqnarray}
n_Q \equiv \sum_{i=1}^3 (n_i - \overline n_i) = \frac{1}{6\pi^2}
\sum_{i=1}^3 [\pi^2 \mu_i T_\nu^2 + \mu_i^3]
	\end{eqnarray}
where $\overline n_i$ denotes the number density of
antineutrinos. Thus, if the neutrinos have a charge $q$, this
will produce a net charge density of the magnitude $qn_Q$. The
argument leading to Eq.\ (\ref{nq}) now gives
	\begin{eqnarray}
\frac{q}{e}
\sum_{i=1}^3 [\pi^2 \mu_i T_\nu^2 + \mu_i^3] \leq 4.8 \times
10^{-25} h_0^2 \; {\rm cm}^{-3} \,,
	\end{eqnarray}
where $e=\sqrt{\alpha}$ in our natural units where $\hbar=c=1$.
This is a strong correlated limit on the possible values of neutrino
charge and degeneracy. In the specific case when the neutrinos
are maximally degenerate as given by Eq.\ (\ref{50}), we obtain
the following bound on neutrino charge:
	\begin{eqnarray}
|q| \leq 5.2 \times 10^{-33} e \times \sqrt{h_0} \,.
	\end{eqnarray}
This is more stringent than any other limit derived so far. The
derivation of the above limit assumes complete degeneracy of
neutrino with zero rest mass. However if the chemical potential
of neutrino is zero then there would be no net charge excess and
the above argument is not applicable.

In conclusion, we would like to point out that the  constraints
we derived here are direct consequences of the observed
anisotropy in cosmic microwave background radiation, and hence
they are more stringent than the other limits derived earlier.
Our limit from the anisotropy of CMBR implies that it is very
unlikely to have any other profound consequences due to the
electric charge asymmetry of the Universe or the electric charge
of neutrinos since such consequences require much higher
asymmetries than are allowed by our limit.

We thank C. Sivaram for discussions.

\end{document}